\documentclass[12pt]{article}

\usepackage{epsfig,graphics}
\usepackage{amssymb}

\newcommand{\beqn}{\begin{eqnarray}}
\newcommand{\eeqn}{\end{eqnarray}}
\newcommand{\be}{\begin{equation}}
\newcommand{\ee}{\end{equation}}
\newcommand{\ba}{\begin{array}}
\newcommand{\ea}{\end{array}}
\newcommand{\R}{{\rm\bf R}}
\newcommand{\C}{{\rm\bf C}}
\newcommand{\pa}{\partial}

\newcommand{\re}{\ref}
\newcommand{\ci}{\cite}
\newcommand{\la}{\label}
\newcommand{\bfr}{\begin{flushright}}
\newcommand{\efr}{\end{flushright}}
\newcommand{\bfl}{\begin{flushleft}}
\newcommand{\efl}{\end{flushleft}}
\newcommand{\fr}{\frac}

\newcommand{\ti}{\tilde}
\newcommand{\st}{\stackrel}

\newcommand{\ds}{\displaystyle}

\newcommand{\cA}{{\cal A}}

\newcommand{\ve}{\varepsilon}
\newcommand{\vp}{\varphi}
\newcommand{\we}{\wedge}
\newcommand{\de}{\delta}\newcommand{\De}{\Delta}

\newcommand{\om}{\omega}

\newcommand{\na}{\nabla}

\newcommand{\Br}{|\kern-.25em|\kern-.25em|}
\newcommand{\brr}{{|\kern-.15em|\kern-.15em|\kern-.15em}\,}
\newcommand{\ddd}{\st{.\kern-.07em.\kern-.07em.}}
\def\N{{\rm I\kern-.1567em N}}                              
\def\R{{\rm I\kern-.1567em R}}                              
\def\C{{\rm C\kern-4.7pt                                    
\vrule height 7.7pt width 0.4pt depth -0.5pt \phantom {.}}}
\def\Z  {{\sf Z\kern-4.5pt Z}}                              

\begin{document}

\renewcommand{\theequation}{\thesection.\arabic{equation}}
\newtheorem{theorem}{Theorem}[section]
\renewcommand{\thetheorem}{\arabic{section}.\arabic{theorem}}
\newtheorem{definition}[theorem]{Definition}
\newtheorem{deflem}[theorem]{Definition and Lemma}
\newtheorem{lemma}[theorem]{Lemma}
\newtheorem{example}[theorem]{Example}
\newtheorem{remark}[theorem]{Remark}
\newtheorem{remarks}[theorem]{Remarks}
\newtheorem{cor}[theorem]{Corollary}
\newtheorem{pro}[theorem]{Proposition}

\newcommand{\bd}{\begin{definition}}
\newcommand{\ed}{\end{definition}}
\newcommand{\bt}{\begin{theorem}}
\newcommand{\et}{\end{theorem}}
\newcommand{\bqt}{\begin{qtheorem}}
\newcommand{\eqt}{\end{qtheorem}}

\newcommand{\bp}{\begin{pro}}
\newcommand{\ep}{\end{pro}}

\newcommand{\bl}{\begin{lemma}}
\newcommand{\el}{\end{lemma}}
\newcommand{\bc}{\begin{cor}}
\newcommand{\ec}{\end{cor}}

\newcommand{\bex}{\begin{example}}
\newcommand{\eex}{\end{example}}
\newcommand{\bexs}{\begin{examples}}
\newcommand{\eexs}{\end{examples}}

\newcommand{\bexe}{\begin{exercice}}
\newcommand{\eexe}{\end{exercice}}

\newcommand{\br}{\begin{remark} }
\newcommand{\er}{\end{remark}}
\newcommand{\brs}{\begin{remarks}}
\newcommand{\ers}{\end{remarks}}

\newcommand{\pru}{{\bf Proof~~}}


\begin{titlepage}

\begin{center}
{\Large\bf
On Lagrangian Theory for Rotating Charge\\
~\\
Coupled to
the
Maxwell Field
\\

}

\vspace{2cm}

{\large Valeriy Imaykin} \footnote{Supported partly by grants DFG
436 RUS 113/929/0-1 and RFBR 10-01-00578-a}

\medskip

{\it Zentrum Mathematik,
TU M\"{u}nchen\\ Boltzmannstr. 3,
Garching, 85747 Germany}\\
email: ivm61@mail.ru

\medskip

\bigskip

{\large Alexander Komech}$^1\,$ \footnote{
Supported partly by the Alexander von Humboldt
Research Award, and by the Austrian Science Fund (FWF): P22198-N13.}

\medskip

{\it Faculty of Mathematics of
Vienna University\\
and IITP RAS (Moscow)
}\\

email: alexander.komech@univie.ac.at

\medskip

\bigskip

{\large Herbert Spohn}

\medskip

{\it Zentrum Mathematik,
TU M\"{u}nchen\\ Boltzmannstr. 3,
Garching, 85747 Germany}\\

email: spohn@ma.tum.de

\end{center}

\vspace{2cm}


\begin{abstract}
We justify the Hamilton least action principle for the Maxwell-Lorentz
equations with
Abraham's rotating extended electron.
The main novelty in the proof is application of  the variational Poincar\'e equations
on the Lie group $SO(3)$.
The variational equations allow to
derive the corresponding conservation laws
from general  N\"other theory of invariants.

\end{abstract}

\end{titlepage}


\setcounter{equation}{0}

\section{Introduction}

We justify the Hamilton least action principle for the system of
Maxwell-Lorentz
equations with a rotating charged particle. Our main contribution
is the variational
derivation of the {\it Lorentz torque equation}, see equation
(\re{lt0}) below.

First recall the case of a finite system of material points
$(q_i,m_i)$. The {\it
angular momentum} is defined by
\be\la{am}
M:=\sum q_i\we p_i:=\sum q_i\we m_i\dot
q_i.
\ee
By the second and the third Newton laws this implies
\be\la{dotam} \dot
M=\sum q_i\we \dot p_i=\sum q_i\we F_i=\sum q_i\we F_i^{ext}=T,
\ee
where $T$ is
called the {\it external force torque}.
Our aim is to derive the similar {\it torque equation} for a charged rigid body
in the Maxwell
field:
\be\la{lt0}
I\dot\om=e\int(x-q)\we\left[E+E^{ext}+(q+\om\we (x-q))\we(B+B^{ext})
\right]\rho(x-q)dx,
\ee
where $I$ is the moment of inertia, $\om$ is the angular velocity,
and
$\rho(x)$ is a charge distribution, and the right hand side is the
torque of the Lorentz force.

Formally the rigid body can be considered as an infinite system
of material points.
However equation (\re{lt0}) cannot be obtained directly from
(\re{dotam}), since we cannot
correctly take into account all the forces of mutual interaction
between the
different pieces of the rigid body. That is why we look for a
different approach to the
derivation of (\re{lt0}). We show that (\re{lt0}) follows from
the
Hamilton variational least action principle with the standard
interaction term
$-A_0\rho+ \vec A\cdot\vec j$
in the Lagrangian
density.

For the free rigid body $(E,E^{ext},B,B^{ext}=0)$ equation
(\re{lt0}) reduces to the Euler's equations which have been obtained
from the variational
principle first by Poincar\'e \ci{Poin}, and in \ci{AKN} for an
external force field with an axial symmetry.

Note that in our case the fields $E$ and $B$ are generated by the
motion of the
charged body, and $E^{ext}$, $B^{ext}$ are external fields.

Let us comment on previous works. Equation (\re{lt0})
is well recognized since the Abraham's works \ci{Abr2, Abr}.
In \ci[Section 11]{Abr2} Abraham
computed
the Lagrangian
as integral of $-A_0\rho+ \vec A\cdot\vec j$
for standing rotating spherically
symmetric electron subject to external fields obeying very
special symmetry conditions. In this case the Lagrangian
depends only on one variable $\om$, the angular velocity.
However, derivation of the torque equation (\re{lt0}) from the
variational
{it Hamilton
least action principle} remained an open question.

Nodvik applied the variational Hamilton principle
to to the rotating charge in the Euler angles \ci{Nod}.
He deduced the dynamical equation \ci[(5.46)]{Nod} for $\om(t)$
and have established the corresponding conservation laws.
However,
the Nodvik equation
looks
differently from (\re{lt0}). Appel and Kiessling \ci{AK}
suggested a transformation of \ci[(5.46)]{Nod} to (\re{lt0}),
but they did not give the details, see \ci[A.1.4]{AK}.
The direct derivation of the conservation laws
from (\re{lt0})
is presented by Kiessling
in \ci{Kiess}.

We propose an invariant derivation of (\re{lt0}) from the
Hamilton least action principle relying on Poincar\'e equations \ci{Poin,AKN}
on the Lie group $SO(3)$.
We also
deduce the corresponding conservation laws by general
Lagrangian formalism using the N\"other theory of invariants.

The plan of our article is as follows.
In Section 2 we state the
Maxwell-Lorentz equations for the rotating charge,
and introduce the corresponding Lagrangian functional.
Further we deduce the equations from the Hamilton
variational principle relying on the Poincar\'e theory \ci{Poin}
presented in the Appendix A.
In  Section 3 we prove the conservation laws.
Finally, in Appendix B we identify the commutators
of invariant vector fields.


\setcounter{equation}{0}

\section{Maxwell-Lorentz equations}

The Maxwell field consists of the electric field $E(x,t)$ and the
magnetic field
$B(x,t)$ generated by a motion of a rotating charge. The external
fields $E^{ext}$
and $B^{ext}$ are also generated by the corresponding external
charges and currents.
Let the rotating charge be centered at the position $q$ with the
velocity $\dot q$.
For simplicity we assume that the mass distribution, $m\,\rho(x)$,
and the charge
distribution, $e\,\rho(x)$, are proportional to each other.
Here $m$ is the total
mass, $e$ is the total charge, and we use a system of units such
that $m=1$, $e=1$.
The coupling function $\rho(x)$ is a sufficiently smooth radially
symmetric function
of fast decay as $|x|\to\infty$,
$$\rho(x)= \rho_r(|x|).\eqno{(C)}
$$
\subsection{Angular velocity}
Let us denote by $\om(t)\in\R^3$ the angular velocity ``in space'' (in
the terminology of \ci{AKN}) of the charge. Namely, let us fix a
``center'' point $O$ of the rigid body. Then the trajectory of each
fixed point of the body is described by
$$
x(t)=q(t)+R(t)(x(0)-q(0)),
$$
where $q(t)$ is the position of $O$ at the time $t$, and $R(t)\in
SO(3)$. Respectively, the velocity reads
\be\la{dotx}
\dot x(t)=\dot q(t)+\dot R(t)(x(0)-q(0))=\dot q(t)+\dot
R(t)R^{-1}(t)(x(t)-q(t))=\dot q(t)+\om(t)\we(x(t)-q(t)),
\ee
where $\om(t)\in\R^3$ correponds to the skew-symmetric matrix $\dot
R(t)R^{-1}(t)$ by the rule
\be\la{mv}
\dot
R(t)R^{-1}(t)={\cal J}\om(t):=\left(
\ba{ccc}
0 & -\om_3(t) & \om_2(t) \\
\om_3(t) & 0 & -\om_1(t) \\
-\om_2(t) & \om_1(t) & 0
\ea
\right).
\ee
We assume that $x$ and $q$ refer to a certain Euclidean coordinate
system in $\R^3$, and the vector product $\we$ is defined in this
system by standard formulas. The identification (\re{mv}) of a
 skew-symmetric matrix and the corresponding angular velocity
vector is true in any Euclidean coordinate system of the same
orientation as the initial one.

\subsection{Dynamical equations}
Then the system of
Maxwell-Lorentz equations with spin reads, see \ci{Sp}
\be \la{mls}
\dot E=\na\we B-(\dot
q+\om\we(x-q))\rho(x-q),\quad\dot B= - \na\we E,
\ee
\be\la{div} \na\cdot E(x,t)=
\rho (x-q(t)),\,\,\,\,\na\cdot B(x,t)= 0,
\ee
\be\la{lf} \ddot q=\int\, [E+E^{ext}+(\dot
q+\om\we(x-q))\we(B+B^{ext})]\rho(x-q) dx,
\ee
\be\la{lt} I\,\dot \om= \int \,
(x-q)\we[E+E^{ext}+(\dot q+\om\we(x-q))\we(B+B^{ext})]\rho(x-q)dx,
\ee
where $I$ is the moment of
inertia defined by \be\la{ib} I=\fr 23 \int \, x^2\rho(x)dx.
\ee Here the
equations (\re{mls}) are Maxwell equations with the corresponding
charge density and
current, equations (\re{div}) are constraints. The back reaction
of the field onto
the particle is given through the Lorentz force equation (\re{lf}),
and the Lorentz
torque equation (\re{lt}) deals with rotational degrees of freedom.


\subsection{Lagrangian functional and variational principle}

Our main goal is to deduce equations (\re{mls})-(\re{lt}) from the
Hamilton least action principle. First let us introduce
{\it electromagnetic potentials} ${\cal A}=(A_0,A)$,
${\cal A}^{ext}=(A_0^{ext},A^{ext})$:
\be\la{pot} B=\na\we
A,\,\,\,E=-\na A_0-\dot A. \ee
\be\la{potext} B^{ext}=\na\we
A^{ext},\,\,\,E^{ext}=-\na A_0^{ext}-\dot A^{ext}. \ee
Next we define the Lagrangian
\beqn L({\cal A},q,R,  \dot{\cal A}, \dot q,\dot R) &
=\ds\fr{1}{2}\int\left(E^2-B^2\right)dx+\fr{1}{2}\dot
q^2+\fr12
I\om^2-\ds\int[A_0+A_0^{ext}]\rho(x-q)dx+\nonumber \\
 & \ds\int(\dot q+\om\we(x-q))\cdot
[A+A^{ext}]\rho(x-q)dx,\la{Lagrom}
\eeqn
where $E$, $B$ are expressed in terms of $\cA$, $\dot\cA$
by (\re{pot}),
and $\om={\cal J}^{-1}\dot R R^{-1}$
by (\re{mv}).

The last two
integrals represent the interaction term
$$
\int[(A_0+A_0^{ext}]\rho-j[A+A^{ext}]dx
$$
in view of (\re{dotx}).
The corresponding action
functional has the form
\be\la{act}
S=
S(\cA,q,R):=
\int_{t_1}^{t_2} L({\cal A}(t),q(t),R(t),
\dot{\cal A}(t),\dot q(t),\dot R(t))dt
\ee
Then the
Hamilton least action principle reads
\be\la{lap}
\de S(\cA,q,R)=0,
\ee where the
variation is taken over ${\cal A}(t),q(t)$, $R(t)$
with the boundary conditions
\be\la{varclassbound}(\de{\cal A},\de q,\de R)
\vert_{t=t_1}=(\de{\cal A},\de q,\de R)\vert_{t=t_2}=0.
\ee
We assume that all the involved functions and fields are
sufficiently smooth and have
(with all the necessary derivatives) a sufficient decay
as $|x|\to\infty$ so that
partial integrations below could be possible.

Our main result is the following theorem.
\begin{theorem}
The Maxwell-Lorentz system with spin (\re{mls}) to (\re{lt})
is equivalent to the
least action principle (\re{lap})--(\re{varclassbound}).
\end{theorem}
 We will analyze the variations in ${\cal A}$, $q$, $R$ separately,
namely, we prove that \be\la{lapp} \fr{\de S}{\de
{\cal A}}=0\,\,\,\quad(a),\,\,\,\,\,\,\,\,\,\qquad\fr{\de S}{\de q}=0
\,\,\,\quad(b),\,\,\,\,\,\,\,\,\,\qquad\fr{\de S}{\de R}=0
\,\,\,\quad(c) \ee is
equivalent to (\re{mls})--(\re{lt}).


\subsection{Equations for fields and particle trajectory}
{\bf Equations for fields } It is well-known \ci{G} that (\re{lapp}) (a) is equivalent to
\be\la{ELA}
\fr{d}{dt}\fr{\de L}{\de\dot{\cal A}}=L_{\cal A}
\ee
and that the last Euler-Lagrange equations
are equivalent to the Maxwell equations (\re{mls})
with the constraints (\re{div}).

\smallskip

{\bf Remark } Note that the terms with $A_0^{ext}$ and $A^{ext}$
in (\re{Lagrom}) are additive and
remain additive while one makes variations in $q$, and $R$. Then, for
simplicity of exposition, we put in all of the computations below
$A_0^{ext}=0$ and $A^{ext}=0$.

\smallskip

{\bf Equations for particle trajectory }
Similarly, (\re{lapp}) (b) is equivalent to
\be\la{Lqq}
\fr{d}{dt}L_{\dot q}=L_q.
\ee
It remains to check that the last Euler-Lagrange equations
are equivalent to the
Lorentz force equation (\re{lf}).
Let us change variables in the last two integrals of (\re{Lagrom})
 and obtain
$$
\ds-\int A_0(x+q,t)\rho(x)+\int(\dot q+\om\we x)\cdot A(x+q,t)\rho(x)dx.
$$
We have
$$L_{\dot q}=\dot
q+\ds\int A(x+q,t)\rho(x) dx,\,\,\,\,L_q=\ds-\int\na A_0(x+q,t)\rho(x)dx+\int(\dot q+\om\we x)\cdot\na
A(x+q,t)\rho(x) dx.
$$
Here, for a vector field $b(x)$
we denote $b\cdot\na A=\sum b_j\na A_j$. Then the
Euler-Lagrange equation (\re{Lqq}) reads
$$\ddot q+\int\left(\dot A(x+q,t)+ (\dot q\cdot\na)
A(x+q,t)
\right)\rho dx
=-\int\na A_0(x+q,t)\rho(x)dx+\int(\dot
q+\om\we x)\cdot\na
A(x+q,t)
\rho dx.
$$
Substituting
$\dot A(x+q,t)=-E(x+q,t)-\na A_0(x+q,t)$ we
obtain
\be\la{ddq} \ddot q=\int E(x+q,t)\rho(x) dx+
\int\left(\dot
q\cdot\na A(x+q,t)-(\dot q\cdot\na) A(x+q,t)\right)\rho dx+
\int(\om\we x)\cdot\na A(x+q,t)\rho dx.
\ee
To make the notations shorter let us omit the dependence on $t$ in the further computations.
By the identity
$\dot q\cdot\na A(x+q)-(\dot q\cdot\na)A(x+q)=\dot q\we\na\we A(x+q)$
we obtain
\be\la{mlqq}
\int(\dot q\cdot\na A(x+q)-\dot q\cdot\na A(x+q))\rho dx=
\int\dot q\we\na\we A(x+q)\rho dx.
\ee
It remains to check that \be\la{intom}
\int(\om\we
x)\cdot\na A(x+q)\rho dx=\int\om\we x\we\na \we A(x+q)\rho dx.
\ee
Let us check
for the first component, for the rest the computation is similar.
The first component
of the LHS of (\re{intom}) equals
$$
\int\left[(\om_2x_3-\om_3x_2)\pa_1A_1(x+q)+(\om_3x_1-\om_1x_3)\pa_1A_2(x+q)+
(\om_1x_2-\om_2x_1)\pa_1A_3(x+q)\right]\rho dx.
$$
The first component of the RHS of (\re{intom}) equals
$$
\int\left[(\om_3x_1-\om_1x_3)(\pa_1A_2(x+q)-\pa_2A_1(x+q))-
(\om_1x_2-\om_2x_1)(\pa_3A_1(x+q)-\pa_1A_3(x+q))\right]\rho dx.
$$
For the difference of the LHS and the RHS we apply partial integration,
and obtain
$$
\int\left[(\om_2x_3-\om_3x_2)\pa_1A_1(x+q)+(\om_3x_1-\om_1x_3)\pa_2A_1(x+q)+
(\om_1x_2-\om_2x_1)\pa_3A_1(x+q)\right]\rho dx=
$$
$$
-\int A_1(x+q)\left[(\om_2x_3-\om_3x_2)\pa_1+(\om_3x_1-\om_1x_3)\pa_2+
(\om_1x_2-\om_2x_1)\pa_3\right]\rho dx=
$$
$$
-\int A_1(x+q)\left[\om_1(x_2\pa_3-x_3\pa_2)+\om_2(x_3\pa_1-x_1\pa_3)+
\om_3(x_1\pa_2-x_2\pa_1)\right]\rho dx=
$$
$$
-\int A_1(x+q)(\om\cdot\na_{\theta})\rho dx,
$$
where $\na_\theta=(\na_{\theta_1}, \na_{\theta_2}, \na_{\theta_3})$
and $\na_{\theta_j}$ is the differentiation in the angular coordinate
$\theta_j$
around the coordinate axis $x_j$. Since $\rho$ is radially symmetric,
one has
$\na_{\theta}\rho=0$ and we come to (\re{intom}). From (\re{pot}),
(\re{mlqq}), and
(\re{intom}) we conclude that the equation (\re{ddq}) reads (\re{lf}).


\subsection{The torque equation}
Finally, it remains to check that (\re{lapp}) (c) is
equivalent to
(\re{lt}).
\medskip\\
To make the corresponding variation in $R$,
let us express
$\om(t)=J^{-1}\dot R(t) R^{-1}(t)$
in the right-invariant vector fields on $SO(3)$.
Namely,
consider an orthonormal basis $\{e_k\}$ with the right orientation in
$\R^3$. Then
\be\la{basis}
e_1\we e_2=e_3,\,\,\,e_3\we e_1=e_2,\,\,\,e_2\we e_3=e_1.
\ee
Let us express the angular velocity in $\{e_k\}$:
$\om(t)=\sum\om_k(t)e_k$. The space $so(3)$ of skew-symmetric $3\times3$
matrices
with the matrix commutator is
isomorphic to $\R^3$
with vector product by the isomorphism ${\cal J}$ of (\re{mv}):
\be\la{isomv}
\left(
\ba{ccc}
0 & -\om_3 & \om_2 \\
\om_3 & 0 & -\om_1 \\
-\om_2 & \om_1 & 0
\ea
\right)={\cal J}(\om_1,\om_2,\om_3).
\ee
In detail, if $A,B\in so(3)$, $a,b\in\R^3$, and
$A={\cal J}a$, $B={\cal J}b$ by the isomorphism (\re{isomv}),
then

\be\la{ABBA}
AB-BA={\cal J}(a\we b).
\ee

Further, $\dot R(t)R^{-1}(t)\in T_E SO(3)$ is the
tangent vector $\dot R(t)$ of $SO(3)$ at the point $R(t)$ translated
to the unit $E$ of $SO(3)$ by the right
translation $R^{-1}(t)$.
By the linear isomorphism (\re{isomv}),
\be\la{dRR}
\dot R(t)R^{-1}(t)=\sum\om_k(t)\ti
e_k,~~~~~~~~~ \ti e_k:={\cal J}^{-1}e_k.
\ee
%
%
%
%
%
%
%

Then
\be\la{dR}
\dot
R(t)=\dot R(t)R^{-1}(t)R(t)=\sum\om_k(t)v_k(R(t)),~~~~~~~~~
v_k(R):=\ti e_k R.
\ee

As the result,
$\dot R(t)$ has the same coordinates w.r.t. the vector fields $v_k$ at
the point $R(t)$ as $\om(t)$ in the basis
$\{e_k\}$.
The fields
$v_k(R)$ are right translations of $\ti e_k$ and hence are right-invariant.

%
%
The next lemma is proved in Appendix B.
\begin{lemma}\la{reffields}
For the
constructed above vector fields $v_k$ on $SO(3)$ the following commutation
relations hold: \be\la{cr2}
[v_1,v_2]=-v_3,\,\,\,[v_3,v_1]=-v_2,\,\,\,[v_2,v_3]=-v_1. \ee
\end{lemma}

{\bf Poincar\'e equations }
Now we are going to deduce (\re{lt}) from the
{\it Poincar\'e equations}.
Namely,
as shown by Poincar\'e \ci{Poin,AKN}, the equation (\re{lapp}) (c) is
equivalent to
the Poincar\'e equations
\be\la{PEE} \fr{d}{dt}\fr{\pa\hat
L}{\pa\om_k}=\sum_{ij}c^j_{ik}\om_i\fr{\pa\hat L}{\pa\om_j}+v_k(\hat
L),\,\,\,k=1,2,3, \ee where $\hat L$ is $L$ with $\om(t)$ expressed in the
coordinates $(\om_1(t),\om_2(t),\om_3(t))$.
For convenience of the reader the derivation of Poincar\'e equations is
presented in
Appendix A.

Note that the Lagrangian $\hat L$ does not
depend explicitly on $R$, and hence $v_k(\hat L)=0$, $k=1,2,3$ by (\re{Lom}).
Hence,
the corresponding Poincar\'e equation reads
$$
\fr{d}{dt}\fr{\pa\hat L}{\pa\om_1}
=\sum_{ij}c^j_{i1}\om_i\fr{\pa\hat L}{\pa\om_j}.
$$
\medskip\\
{\bf Equivalence to the torque equation }
It remains to check that the equations (\re{PEE}) are equivalent to the
Lorentz torque equation (\re{lt}).
It suffices to check the equivalence for the first component with $k=1$,
since for
the rest $k$ the computation is similar.
\medskip\\
First,
$$
\fr{\pa\hat L}{\pa\om_1}=I\om_1+\int(x_2A_3(x+q)-x_3A_2(x+q))\rho(x)dx.
$$
Therefore,
$$
\fr{d}{dt}\fr{\pa\hat L}{\pa\om_1}=\int(x_2(\dot A_3(x+q)+\dot
q\cdot\na A_3(x+q))-
x_3(\dot A_2(x+q)+\dot q\cdot\na A_2(x+q)))\rho\, dx.
$$
Second, by (\re{cr2}) and (\re{const}) one has $c^2_{31}=-1$, $c^3_{21}=1$,
 and all the rest $c^j_{i1}$=0.
Hence,
$$
\sum_{ij}c^j_{i1}\om_i\fr{\pa\hat L}{\pa\om_j}=\om_2\fr{\pa\hat
L}{\pa\om_3}-\om_3\fr{\pa\hat L}{\pa\om_2}=
$$
$$
\om_2\left[I\om_3+\int(x_1A_2(x+q)-x_2A_1(x+q))\rho
dx\right]-\om_3\left[I\om_3+\int(x_3A_1(x+q)-x_1A_3(x+q))\rho dx\right]=
$$
$$
\int\left[-(\om_2x_2+\om_3x_3)A_1(x+q)+x_1(\om_2A_2(x+q)+\om_3A_3(x+q))
\right]\rho\,
dx.
$$
Finally, we come to the equation
$$
I\dot\om_1=\int(x_3\dot A_2(x+q)-x_2\dot A_3(x+q))\rho\,dx+
\int\left[x_3(\dot
q\cdot\na)A_2(x+q)-x_2(\dot q\cdot\na)A_3(x+q)\right]\rho\,dx+
$$
\be\la{P1}
\int\left[x_1(\om_2A_2(x+q)+\om_3A_3(x+q))-(\om_2x_2+
\om_3x_3)A_1(x+q)\right]\rho\,
dx. \ee

Now let us proceed to the first component of the equation (\re{lt}).
Using the identity
$$
x\we[(\om\we x)\we B]=(\om\we x)(x\cdot B)
$$
we obtain
\be\la{lt1}
I\dot\om_1=\int(x\we E(x+q))_1\rho\,dx+\int(x\we
(\dot q\we
B(x+q)))_1\rho\,dx+\int(\om\we x)_1(x\cdot B(x+q))\rho\,dx,
\ee
We insert $B=\na\we
A$ and obtain that the first integral equals
$$
\int\left[x_3\dot A_2(x+q)-x_2\dot A_3(x+q)+(x_3\pa_2-x_2\pa_3)\vp\right]
\rho\,dx=
$$
$$
\int(x_3\dot A_2(x+q)-x_2\dot A_3(x+q))\rho\,dx-\int\vp(x_3\pa_2-
x_2\pa_3)\rho\,dx=
$$
\be\la{one} \int(x_3\dot A_2(x+q)-x_2\dot A_3(x+q))\rho\,dx,
\ee
since
$(x_3\pa_2-x_2\pa_3)\rho=\pa_{\theta_1}\rho=0$. Further, the
second integral of
(\re{lt1}) reads
$$
\int\left[x_2(\dot q_1(\pa_3A_1(x+q)-\pa_1A_3(x+q))- \dot
q_2(\pa_2A_3(x+q)-\pa_3A_2(x+q))-\right.
$$
\be\la{two}\left.x_3(\dot q_3(\pa_2A_3(x+q)-\pa_3A_2(x+q))- \dot
q_1(\pa_1A_2(x+q)-\pa_2A_1(x+q))\right]\rho\,dx.\ee Finally, the third
integral of
(\re{lt1}) gives
$$
\int\left[x_1(\om_2A_2(x+q)+\om_3A_3(x+q))-(\om_2x_2+\om_3x_3)A_1(x+q)
\right]\rho\,
dx+
$$
$$
\int(\om_2x_3-\om_3x_2)(A_1(x+q)(x_3\pa_2-x_2\pa_3)+A_2(x+q)(x_1\pa_3-x_3
\pa_1)+
A_3(x+q)(x_2\pa_1-x_1\pa_2))\rho\,dx=
$$
\be\la{three}
\int\left[x_1(\om_2A_2(x+q)+\om_3A_3(x+q))-(\om_2x_2+\om_3x_3)A_1(x+q)
\right]\rho\,
dx, \ee since $$(A_1(x+q)(x_3\pa_2-x_2\pa_3)+A_2(x+q)(x_1\pa_3-x_3\pa_1)+
A_3(x+q)(x_2\pa_1-x_1\pa_2))\rho=A(x+q)\cdot\na_{\theta}\rho=0.
$$
By (\re{one}),
(\re{two}), (\re{three}), the difference between the RHS of (\re{P1}) and
the RHS of
(\re{lt1}) equals
$$
\int\left[x_2(\dot q_1\pa_3A_1(x+q)+\dot q_2\pa_3A_2(x+q))- x_3(\dot
q_3\pa_2A_3(x+q)+\dot q_1\pa_2A_1(x+q))\right]\rho\,dx+
$$
$$
\int\left[x_2\dot q_3\pa_3A_3(x+q)-x_3\dot q_2\pa_2A_2(x+q)\right]\rho\,dx=
$$
$$
\int\dot q\cdot A(x+q)(x_3\pa_2-x_2\pa_3)\rho\,dx=0
$$
and we obtain that the equation (\re{P1}) reads (\re{lt1}).
The theorem is proved.
\hfill$\Box$


\setcounter{equation}{0}

\section{Conservation laws}

We have derived the system (\re{mls})-(\re{lt}) by the least action
principle with
the Lagrangian (\re{Lagrom}). When the external fields
posess a symmetry,
the correponding
conservation laws can be also derived from the Lagrangian formalism.
Let us
recall that the Lagrangian (\re{Lagrom})
reads
\beqn
L({\cal A},q,R,  \dot{\cal A}, \dot q,\dot R)=\ds\fr{1}{2}\int\left(E^2-B^2
\right)dx+\fr{1}{2}\dot q^2+\fr12I\om^2-\nonumber\\
 \nonumber\\
\ds\int (A_0+A_0^{ext})\rho(x-q)dx+\!\ds\int(\dot q+\om\we(x-q))
\cdot (A+A^{ext})\rho(x-q)dx,\la{Lagrom0}
\eeqn
where $\om={\cal J}^{-1}\dot RR^{-1}$, ${\cal A}=(A_0,A)$,
$\dot {\cal A}=(\dot A_0,\dot A)$, and
\be\la{pot2} E=-\na A_0-\dot A,\,\,\,\, B=\na\we A.\ee
As above, we denote
\be \la{Lagrom1}
\hat L({\cal A},q,R,  \dot{\cal A}, \dot q,\om)=
L({\cal A},q,R,  \dot{\cal A}, \dot q,\dot R)
\ee
where $\om=(\om_1,\om_2,\om_3)$
is defined by (\re{dR}) i.e.
$\om_k$
are coordinates of
$\dot R$ in the basis $v_1(R),v_2(R),v_3(R)$.



\subsection{Energy conservation}
Let us denote
\be\la{XV}
X:=({\cal A},q,R),\,\,\,V:=\dot X=(\dot{\cal A},\dot
q,\dot R).
\ee
Let $A_0^{ext}$ end $A^{ext}$ do not depend  on time.
Then the Lagrangian (\re{Lagrom0}) does not depend on $t$, and the {\it energy}
\be\la{en}
E(X,V):=L_V\cdot V-L
\ee
is conserved, \ci{Arn}. By (\re{XV}) and since $L$ does
not depend on $\dot A_0$, we have

\be\la{LVV}
L_V\cdot V=L_{\dot A}\cdot\dot A+L_{\dot
q}\cdot\dot q+L_{\dot R}\cdot\dot R=
\hat L_{\dot A}\cdot\dot A+\hat L_{\dot q}\cdot\dot q+\hat
L_{\om}\cdot\om.
\ee

\begin{pro}\la{Econ}
The energy  reads
\be\la{energy}
E=\fr12\int(|E|^2+|B|^2)dx+\fr12\dot q^2+\fr12 I\om^2+\int\,A_0^{ext}\rho(x-q)dx.
\ee
\end{pro}
{\bf Proof } By (\re{Lagrom1}) and (\re{Lagrom0}),
one has
$$
\hat L_{\dot A}\cdot\dot A=-\int\,E\cdot\dot A\,dx,\,\,\,
\hat L_{\dot q}\cdot\dot q=\dot q^2+\int\,\dot q\cdot(A+A^{ext})\rho(x-q)dx,
$$
and
$$
\hat L_{\om}\cdot\om=I\om^2+\int\,\om\we(x-q)\cdot(A+A^{ext})\rho(x-q)dx.
$$
Then
$$
E=\hat L_{\dot A}\cdot\dot A+\hat L_{\dot q}\cdot\dot q+\hat
L_{\om}\cdot\om-\hat L
$$

\be\la{en1}
=\fr12\dot q^2+\fr12 I\om^2+\fr12\int(|B|^2-|E|^2)dx+
\int(-E\cdot\dot A+A_0\cdot(\na\cdot E)+A_0^{ext}\rho(x-q))dx.
\ee
The last integral equals
$$
-\int(E\cdot\dot A+\na A_0\cdot E+A_0^{ext}\rho(x-q))dx=-\int\,(E(\dot A+\na A_0)+A_0^{ext}\rho(x-q))dx
$$
$$
=
\int\,E^2dx+\int\,A_0^{ext}\rho(x-q)dx
$$
and hence (\re{en1}) reads (\re{energy}).\hfill$\Box$


\subsection{Momentum conservation}

Let us consider the spatial translations of the lagrangian coordinate $X=({\cal A},q,R)$:
$$
({\cal A}(x),q,R)\mapsto({\cal A}(x-h),q+h,R).
$$
If the external field ${\cal A}^{ext}=(A_0^{ext},A^{ext})$
does not depend on $x_j$
with some $j$,
then the Lagrangian (\re{Lagrom0}) is invariant w.r.t to the one-parametric group of spatial translations
\be\la{gsX}
g_s^j({\cal A}(x),q,R)=({\cal A}(x-se_j),q+se_j,R),
\ee
where $e_j\in\R^3$ is the corresponding basis vector.
By the  N\"other theorem \ci{Arn} the expression
\be\la{mom}
P_j=P_j(X,V):=L_V\cdot\fr{dg_s^j(X)}{ds}\vert_{s=0}
\ee
is conserved.

\begin{definition}
Vector $P=(P_1,P_2,P_3)$ is called  momentum
of the state $(X,V)$.
\end{definition}
\begin{pro}\la{Mcon}
The momentum reads (cf. \ci{Kiess})
\be\la{PP}
P=\dot q+\int\,E\we B\,dx+\int\,A^{ext}\rho(x-q)dx.
\ee
\end{pro}
{\bf Proof }
For concreteness let us compute $P_1$.
Formula (\re{gsX}) implies
$$
\fr{dg_s^1(X)}{ds}\vert_{s=0}=
-(e_1\cdot\na A(x),\,\,\,e_1,0).
$$
Since $L$ does not depend on $\dot A_0$, and the map $g_s^1$ leaves $\om$
unchanged,
$$
P_1=
L_V\cdot\fr{dg_s^1(X)}{ds}\vert_{s=0}=-L_{\dot A}\cdot(e_1
\cdot\na)A+L_{\dot q}\cdot e_1=
$$
$$
-\int(\na A_0+\dot A)\cdot(e_1\cdot\na)A\,dx+\dot qe_1+\int\,e_1
\cdot A\rho(x-q)dx+\int A^{ext}_1\rho(x-q)dx.
$$
Note that
$$
\dot q_1+\int\,A_1\rho(x-q)dx-\int(\na A_0+\dot A)\cdot\pa_1A\,dx=
$$
\be\la{n1}
\dot q_1+\int\,A_1\rho(x-q)dx-\int(\pa_1A_0\pa_1A_1+\pa_2A_0\pa_1A_2+
\pa_3A_0\pa_1A_3+\dot A_1\pa_1A_1+\dot A_2\pa_1A_2+\dot A_3\pa_1A_3)dx.
\ee
On the other hand,
consider the RHS of (\re{PP}) and obtain
$$
\dot q_1+\int(E\we B)_1dx=\dot q_1+\int[(-\pa_2A_0-\dot A_2)(\pa_1A_2-
\pa_2A_1)+(\pa_3A_0+\dot A_3)(\pa_3A_1-\pa_1A_3)]dx=
$$
\be\la{M1} \dot q_1+\int(-\pa_2A_0\pa_1A_2-\dot A_2\pa_1A_2+
\pa_2A_0\pa_2A_1+\dot
A_2\pa_1A_2+\pa_3A_0\pa_3A_1+\dot A_3\pa_3A_1-\pa_3A_0\pa_1A_3-
\dot A_3\pa_1A_3)dx.
\ee
The difference between (\re{n1}) and (\re{M1}) equals
\be\la{dif}
\int\,A_1\rho(x-q)dx-\int(\pa_1A_0\pa_1A_1+\pa_2A_0\pa_2A_1+
\pa_3A_0\pa_3A_1+\dot
A_1\pa_1A_1+\dot A_2\pa_2A_1+\dot A_3\pa_3A_1)dx.
\ee
Finally,
the partial integration implies
$$
\int\,A_1\rho(x-q)dx=\int\,A_1(\na\cdot E)dx=
\int\,A_1\na\cdot(-\na A_0-\dot A)dx=\int\,A_1(-\De A_0-\na\dot A)dx=
$$
$$
\int(\pa_1A_0\pa_1A_1+\pa_2A_0\pa_2A_1+\pa_3A_0\pa_3A_1+
\dot A_1\pa_1A_1+\dot A_2\pa_2A_1+\dot A_3\pa_3A_1)dx
$$
 Hence, the difference (\re{dif}) equals zero and the first components of the LHS and RHS of (\re{PP}) are equal.
 \hfill$\Box$


\subsection{Angular momentum conservation}

Let the external potential ${\cal A}^{ext}$ be axially
symmetric,
\be\la{Uk}
{\cal A}^{ext}(U_k x)=U_k{\cal A}^{ext}(x),
\ee
where $U_k$ is any rotation around the axis $x_k$.

\begin{lemma}
Let (\re{Uk}) hold.
Then the Lagrangian (\re{Lagrom0}) is invariant w.r.t.
the axial rotations
\be\la{frot} A_0(x)\mapsto
A_0(U_k^{-1}x),\,\,\,A(x)\mapsto U_kA(U_k^{-1}x),\,\,\,\dot A(x)
\mapsto U_k\dot A(U_k^{-1}x);
\ee \be\la{vrot} q\mapsto U_kq,\,\,\,\dot q\mapsto U_k\dot q,
\ee
\be\la{Rrot} R\mapsto U_kR,\,\,\,\,\dot R\mapsto U_k\dot R,
\ee.
\end{lemma}
{\bf Proof } By (\re{pot2}) the transforms (\re{frot}) of the
potentials induce the following transforms of the fields:
\be\la{EBrot}
E(x)\mapsto U_kE(U_k^{-1}x),\,\,\,\,B(x)\mapsto U_kB(U_k^{-1}x).
\ee
Further, we have, in operator notations, ${\cal J}\om=\om\we$,
where $\om\we$ is the operator of the vector product by $\om$ in $\R^3$.
Then it is easy to check that ${\cal J}(U_k\om)=U_k{\cal J}(\om)U_k^{-1}$.
Thus, for $\om={\cal J}^{-1}\dot RR^{-1}$ we obtain ${\cal J}\om
=\dot RR^{-1}$ and hence ${\cal J}(U_k\om)=U_k(\dot RR^{-1})U_k^{-1}
=(U_k\dot R)(U_kR)^{-1}$. Finally,
$$
U_k\om={\cal J}^{-1}(U_k\dot R)(U_kR)^{-1}.
$$
This means that the transforms (\re{Rrot}) induce the following
transform of $\om$: \be\la{omrot}\om\mapsto U_k\om.
\ee
Now it is easy to check, in view of axial symmetry of
${\cal A}^{ext}$, the invariance of
$L$ w.r.t. the transforms (\re{EBrot}), (\re{vrot}),
(\re{omrot})
since $\rho$ is spherical symmetric. \hfill $\Box$
\medskip

Recall that $\ti e_k$ is the preimage of the basis vector
$e_k$ w.r.t. the isomorphism
(\re{isomv}).
The Lagrangian $\hat L$ (\re{Lagrom1}) is invariant w.r.t.
the spatial rotations (\re{EBrot}), (\re{vrot}), (\re{omrot}). In particular, $\hat L$ is invariant
under the transform groups $g_s^k=e^{s\ti e_k}$.
Hence, by the N\"other theorem the expression
\be\la{Mk}
M_k=M_k(X,V):=
\hat L_V\cdot\frac{\ds dg_s(X)}{\ds ds}\vert_{s=0}
\ee
is conserved.

\begin{definition}
Vector $M=(M_1,M_2,M_3)$ is called
angular momentum
of the state $(X,V)$.
\end{definition}

\begin{pro}
The  angular momentum reads
\be\la{MM}
M=q\we\dot q+I\om+\int\,x\we E\we Bdx+\int x\we A^{ext}\rho(x-q)dx.
\ee
\end{pro}
{\bf Proof } For concreteness let us compute $M_k$ with $k=1$.
Then $g_s=e^{s\ti e_1}$, and hence
$$
g_sX=(A_0(e^{-s\ti e_1}x),e^{s\ti e_1}A(e^{-s\ti e_1}x),
e^{s\ti e_1}q,e^{s\ti e_1}).
$$
Then
$$
\fr{d}{ds}g_s(X)\vert_{s=0}=
$$

$$
(-\ti e_1e^{-s\ti e_1}x\cdot\na)A_0(e^{-s\ti e_1}x),
\ti e_1e^{s\ti e_1}A(e^{-s\ti
e_1}x)+e^{s\ti e_1}(-\ti e_1e^{-s\ti e_1}x\cdot\na)A(e^{-s\ti e_1}x),
\ti e_1e^{s\ti
e_1}q,\ti e_1e^{s\ti e_1})\vert_{s=0}
$$

$$
=(\ti e_1A_0(x),\ti e_1A(x)-(\ti e_1x\cdot\na)A(x),\ti e_1q,e_1).
$$
Note that the last component is the coordinates of $\ti e_1$ w.r.t.
the basis $\ti
e_1,\ti e_2, \ti e_3$ and thus equals $e_1=(1,0,0)$. Then, since
$\hat L$ does not
depend on $\dot A_0$,
$$
M_1=\hat L_V\cdot\fr{d}{ds}g_s(X)\vert_{s=0}
=\hat L_{\dot A}\cdot(\ti e_1A(x)-(\ti
e_1x\cdot\na)A(x))+\hat L_{\dot q}\cdot(\ti e_1q)+\hat L_{\om}\cdot e_1
$$
$$
=\int\,dx\left(\dot A\cdot(\ti e_1A(x)-(\ti e_1x\cdot\na)A(x))
+\na A_0\cdot(\ti
e_1A(x)-(\ti e_1x\cdot\na)A(x))\right)
$$

$$
+\dot q\cdot(\ti e_1q)+\int(\ti e_1q)\cdot (A+A^{ext})\rho(x-q)dx+I\om\cdot
e_1+\int(e_1\we(x-q))\cdot[A+A^{ext}]\rho(x-q)dx
$$

$$
=(q\we\dot q)_1+I\om_1+\int(x_2 A^{ext}_3-x_3 A^{ext}_2)\rho(x-q)dx
$$

\be\la{second}
+\int(x_2A_3-x_3A_2)\rho(x-q)dx+\int(\dot A+\na A_0)\cdot((0,-A_3,A_2)
+(x_3\pa_2-x_2\pa_3)A)dx.
\ee

We have to prove that this expression
equals to the first component of the RHS of (\re{MM}).
It suffices
to prove that
the last line
equals to the first component of $\ds \int\,x\we(E\we B)dx$.
Indeed, $\rho(x-q)=\na\cdot E=\na\cdot(-\na A_0-\dot A)$,
hence
$$
\int(x_2A_3-x_3A_2)\rho(x-q)dx=
\int(x_2A_3-x_3A_2)(-\na\dot A-\na^2A_0)dx
$$
\be\la{first}
~~~~~~~~~~~~~~~~~~~~~~~~~~~~~
~~~~~=\int\na(x_2A_3-x_3A_2)(\dot A+\na A_0)dx.
\ee
Then (\re{second}) transforms to
$$
\int\left(\pa_1(x_2A_3-x_3A_2)(\dot A_1+\pa_1A_0)+x_2\pa_2A_3(\dot
A_2+\pa_2A_0)-x_3\pa_3A_2(\dot A_3+\pa_3A_0)\right)dx~~~~~~
$$
\be\la{summa}
+\int\left((x_3\pa_2-x_2\pa_3)A_1(\dot A_1+\pa_1A_0)-x_2\pa_3A_2(\dot
A_2+\pa_2A_0)+ x_3\pa_2A_3(\dot A_3+\pa_3A_0)\right)dx. \ee
On the other hand,
substitute $E=-\dot A-\na A_0$, $B=\na\we A$ and obtain that the first
component of
$\ds\int x\we(E\we B)dx$ equals
$$
~~\int x_2((\pa_1A_3-\pa_3A_1)(\dot A_1+\pa_1A_0)+(\pa_2A_3-\pa_3A_2)(\dot
A_2+\pa_2A_0))dx
$$
$$
-\int x_3((\pa_3A_2-\pa_2A_3)(\dot A_3+\pa_3A_0)+(\pa_1A_2-\pa_2A_1)(\dot
A_1+\pa_1A_0))dx
$$
which coincides with (\re{summa}). The proof is complete.\hfill $\Box$


\appendix


\setcounter{equation}{0}

\section{Poincar\'e equations}

The derivation of Poincar\'e equations is presented for the convenience of the
reader, our exposition follows \ci{AKN}. Poincar\'e has obtained the form of
the Hamilton least action principle for Lagrangian systems on
manifolds \ci{Poin}.

Let $v_1,\dots,v_n$ be vector fields on a $n$-dimensional manifold $M$
which are
linearly independent at every point. Then the commutation
relations hold,
$$[v_i,v_j](g)=\sum c_{ij}^k(g)v_k(g),~~~~~~~g\in M
$$
where the commutator $[v_i,v_j]$ is defined by
$$
[v_i,v_j](f):=v_i(v_j(f))-v_j(v_i(f)),
$$
and $v(f)$ is the derivative of a smooth function $f$ on $M$ w.r.t.
the vector field $v$.

If $g(t)$ is a smooth path in $M$
and $f$ is a smooth function on $M$, one has
$\dot g(t)=\sum\om_i(t)v_i(g(t))$ and
$$
\fr{d}{dt}f(g(t))=f'(g(t))\cdot\dot g=
f'(g(t))\cdot\sum\om_i(t)v_i(g(t))
=\sum
v_i(f)\om_i(t).
$$
Now
consider a variation $g(\ve,t)$ of the path $g(t)$. Then
similarly,
$$
\pa_\ve f(g(\ve,t))=\sum_jv_j(f)w_j(\ve,t),
$$
where $w_j(\ve,t)$ are coordinates of $\fr{\pa g}{\pa\ve}(\ve,t)
\in T_{g(\ve,t)}M$.
Hence
$$
\pa_\ve\pa_t
f(g(\ve,t))=\sum_i\sum_jv_j(v_i(f))w_j\om_i
+\sum_iv_i(f)\om_i',
$$
$$
\pa_t\pa_\ve
f_\ve(g(\ve,t)=\sum_j\sum_iv_i(v_j(f))w_j\om_i
+\sum_jv_j(f)\dot w_j,
$$
where the prime resp. dot
stand for the differentiation in $\ve$ resp. $t$.
However,
the differentiations in $t$ and $\ve$ commute, hence
we obtain
by subtraction
$$
\sum_k v_k(f)\om_k'=\sum_k\sum_{ij}c^k_{ij}\om_iw_jv_k(f)
+\sum_k v_k(f)\dot w_k.
$$
Since $f$ is an arbitrary smooth function, we come to the
 relations
\be\la{cr} \om_k'(\ve,t)=\sum_{ij}c^k_{ij}\om_iw_j+\dot w_k.
\ee

Further, let us consider a Lagrangian function $L(\dot g,g)$ on $TM$.
Then $L(\dot
g,g)$ can be expressed in the variables $\om$: $L(\dot g,g)=\hat L(\om,g)$.
Let us
compute the variation of the corresponding action functional
taking (\re{cr})
into
account:
$$
\fr{d}{d\ve}\int_{t_1}^{t_2}\hat L(\om(\ve,t),g(\ve,t))dt=
\int_{t_1}^{t_2}\left(\sum_k\fr{\pa\hat L}{\pa\om_k}\om_k'+ \na_g\hat L\cdot
g'\right)dt=
$$
$$
\int_{t_1}^{t_2}\left[\sum_k\fr{\pa\hat L}{\pa\om_k}(\dot
w_k+\sum_{ij}c^k_{ij}\om_iw_j) +\na_g\hat L\cdot\sum_kw_kv_k\right]dt=
$$
$$
\sum_k\fr{\pa\hat L}{\pa\om_k}w_k\Big|_{t_1}^{t_2}+\int_{t_1}^{t_2}
\sum_k\left[-\fr{d}{dt}\fr{\pa\hat L}{\pa\om_k}
+\sum_{ij}c^j_{ik}\om_i\fr{\pa\hat
L}{\pa\om_j}+v_k(\hat L)\right]w_k\,dt.
$$

The variation should be zero by the Hamilton least action principle, under the
boundary value conditions
\be\la{star}
g(\ve,t_1)=g_1,\,\,\,g(\ve,t_2)=g_2.
\ee Since
$w_k(t_1)=w_k(t_2)=0$ by (\re{star}), we obtain the following
{\it Poincar\'e
equations}:
\be\la{Pe}
\fr{d}{dt}\fr{\pa\hat
L}{\pa\om_k}=\sum_{ij}c^j_{ik}\om_i\fr{\pa\hat L}{\pa\om_j}+v_k(\hat L).
\ee {\bf
Remarks } 1. If $g$ is expressed in a local map as
$(g_1,...,g_n)\in\R^n$, and $v_k=\pa_{g_k}$, then (\re{Pe}) reduce
to the standard Euler-Lagrange equations.

\smallskip

\noindent 2. If a Lagrangian $L$ does not depend on $g$,
$\hat L=\hat L(\om)$ one has
\be\la{Lom} v_k(\hat L)=0. \ee Indeed, $v_k(\hat L)=\na_g\hat L\cdot v_k(g)=0$.

\smallskip

\noindent 3. Suppose $M=G$ is a Lie group, and  let $v_k$, $k=1, ...,\,n$ be
independent either left-invariant or right-invariant vector fields on $G$.
Then $c^k_{ij}(g)$ are constant:

\be\la{const} c^k_{ij}(g)\equiv c^k_{ij},~~~~~~~~\,g\in G.
\ee

\setcounter{equation}{0}
\section{Commutators of invariant vector fields}
 ~~~
{\it Step 1.}
By (\re{ABBA}) the isomorphism (\re{isomv}) translates
relations (\re{basis}) to
\be\la{ccrr}
[\ti e_1,\ti e_2]=\ti e_3,\,\,\,[\ti e_3,\ti e_1]=\ti e_2,
\,\,\,[\ti e_2,\ti e_3]=\ti e_1
\ee
in the sense of matrix commutator.

\smallskip

{\it Step 2.} Recall that the right-invariant vector fields
$v_k$ on $SO(3)$ are defined by right
translations $v_k(R)=\ti e_kR$, where $R\in SO(3)$.
We should prove (\re{cr2})
in the sense of the commutators of vector fields on the Lie group
$SO(3)$.

\smallskip

Since the fields $v_k$ are right-invariant, it suffices
to check the relations (\re{cr2})
 at the group unit $E$. Let us
compute the derivative of a smooth function $f$ on $SO(3)$ w.r.t. a
right-invariant field $v_A$ such that $v_A(E)=A\in so(3)$. In this case
$v_A(R)=AR$ for $R\in SO(3)$. Consider a smooth path $R_1(t)\in SO(3)$ such
that $R_1(0)=R$, $\dot R_1(0)=AR$. Then
$$
v_A(f)(R)=\fr{d}{dt}f(R_1(t))\vert_{t=0}=\left[f'(R_1(t))\cdot\dot
R_1(t)\right]\vert_{t=0}=f'(R)\cdot AR.
$$
In particular,
\be\la{commut}
v_{[A,B]}(f)(E)=f'(E)\cdot[A,B],
\ee
where $[A,B]=AB-BA$ is the matrix commutator in $so(3)$.

Now let us compute $v_A(v_B(f))(E)$ for a right-invariant field $v_B$ such that
$v_B(E)=B\in so(3)$, $v_B(R)=BR$.
Consider a smooth path $R_2(t)\in SO(3)$ such
that $R_2(0)=E$, $\dot R_2(0)=A$. Then
$$
v_A(v_B(f))(E)=\fr{d}{dt}[f'(R_2(t))\cdot
BR_2(t)]\vert_{t=0}=\fr{d}{dt}f'(R_2(t))\vert_{t=0}\cdot
BR_2(t)\vert_{t=0}+f'(R_2(t))\vert_{t=0}\cdot\fr{d}{dt}BR_2(t)\vert_{t=0}
$$
$$
=\left[f''(R_2(t))\cdot\dot R_2(t)\right]\vert_{t=0}\cdot
BR_2(t)\vert_{t=0}+[f'(R(t))\cdot B\dot R_2(t)]\vert_{t=0}
$$
$$
=(f''(E)\cdot A)\cdot B+f'(E)\cdot BA.
$$
Then, since the form $(f''(E)\cdot A)\cdot B$ is symmetric
w.r.t. matrices $A,B$ one has
$$
[v_A,v_B](f)(E)=v_A(v_B(f))(E)-v_B(v_A(f))(E)=f'(E)\cdot(BA-AB)=
-v_{[A,B]}(f)(E).
$$
by (\re{commut}). Together with (\re{ccrr}) this
completes the proof.



\end{document}